\documentclass[sigconf]{acmart}

\usepackage{graphicx}
\graphicspath{{figs/}{experiments/}}

\usepackage{listings} 

\usepackage{algorithmic}
\usepackage{algorithm}

\usepackage{subfigure}
\usepackage{booktabs}
\usepackage{enumitem}
\usepackage{balance}
\usepackage{footnote}
\usepackage{tablefootnote}
\usepackage{wrapfig}

\AtBeginDocument{%
  \providecommand\BibTeX{{%
    \normalfont B\kern-0.5em{\scshape i\kern-0.25em b}\kern-0.8em\TeX}}}

\setcopyright{acmcopyright}
\copyrightyear{2018}
\acmYear{2018}
\acmDOI{10.1145/1122445.1122456}

\acmConference[Woodstock '18]{Woodstock '18: ACM Symposium on Neural
  Gaze Detection}{June 03--05, 2018}{Woodstock, NY}
\acmBooktitle{Woodstock '18: ACM Symposium on Neural Gaze Detection,
  June 03--05, 2018, Woodstock, NY}
\acmPrice{15.00}
\acmISBN{978-1-4503-XXXX-X/18/06}

\begin{document}

\title{GeoFlink: A Distributed and Scalable Framework for the Real-time Processing of Spatial Streams}

\author{Salman Ahmed Shaikh}
\authornotemark[1]
\email{shaikh.salman@aist.go.jp}
\affiliation{%
  \institution{Artificial Intelligence Research Center\\AIST}
  \streetaddress{2-4-7 Aomi, Koto-ku}
  \city{Tokyo}
  \country{Japan}
}

\author{Komal Mariam}
\email{kmariam.msee17seecs@seecs.edu.pk}
\affiliation{%
  \institution{School of Electrical Engineering and Computer Science\\National University of Sciences and Technology}  
  \city{Islamabad}
  \country{Pakistan}}

\author{Hiroyuki Kitagawa}
\email{kitagawa@cs.tsukuba.ac.jp}
\affiliation{%
  \institution{Center for Computational Sciences\\University of Tsukuba}
  \city{Tsukuba}
  \country{Japan} 
}

\author{Kyoung-Sook Kim}
\email{ks.kim@aist.go.jp}
\affiliation{%
  \institution{Artificial Intelligence Research Center\\AIST}
  \streetaddress{2-4-7 Aomi, Koto-ku}
  \city{Tokyo}
  \country{Japan}  
}


\begin{abstract}
Apache Flink is an open-source system for scalable processing of batch and streaming data. Flink does not natively support efficient processing of spatial data streams, which is a requirement of many applications dealing with spatial data. Besides Flink, other scalable spatial data processing platforms including GeoSpark, Spatial Hadoop, etc. do not support streaming workloads and can only handle static/batch workloads. To fill this gap, we present GeoFlink, which extends Apache Flink to support spatial data types, indexes and continuous queries over spatial data streams. To enable the efficient processing of spatial continuous queries and for the effective data distribution across Flink cluster nodes, a gird-based index is introduced. GeoFlink currently supports spatial range, spatial $k$NN and spatial join queries on point data type. An extensive experimental study on real spatial data streams shows that GeoFlink achieves significantly higher query throughput than ordinary Flink processing.


\end{abstract}




\maketitle

\section{Introduction}
\label{sec:introduction}
With the increase in the use of GPS-enabled devices, spatial data is omnipresent. Many applications require real-time processing of spatial data, for instance, route guidance in disaster evacuation, patients tracking to prevent the spread of serious diseases, etc. Such applications entail real-time processing of millions of tuples per second. Existing spatial data processing frameworks, for instance, PostGIS \cite{PostGIS} and QGIS \cite{QGIS2020} are not scalable to handle such huge data and throughput requirements, while scalable platforms like Apache Spark \cite{SparkStreaming}, Apache Flink \cite{ApacheFlinkDoc}, etc. do not natively support spatial data processing, resulting in increased spatial querying cost. Besides, there exist a few solutions to handle large scale spatial data, for instance Hadoop GIS \cite{HadoopGIS2013}, Spatial Hadoop \cite{Spatialhadoop2015}, GeoSpark \cite{JiaYu2019}, etc. However, they cannot handle real-time spatial streams. To fill this gap, we present GeoFlink, which extends Apache Flink to support distributed and scalable processing of spatial data streams.


\begin{lstlisting}[caption={A GeoFlink (Java) code for spatial join query}, label={code:rangeQuery}]
//Defining dataStream boundaries & creating index
double minX = 115.50, maxX = 117.60, 
                           minY = 39.60, maxY = 41.10;
int gridSize = 100;
UniformGrid uGrid = new UniformGrid(
                    gridSize, minX, maxX, minY, maxY);
//Ordinary point stream
DataStream<Point> S1 = SpatialStream.
                PointStream(oStream, "GeoJSON", uGrid);
//Query point stream
DataStream<Point> S2 = SpatialStream.
                PointStream(qStream, "GeoJSON", uGrid);
//Continous join query 
DataStream<Tuple2<String, String>> joinStream =
                     JoinQuery.SpatialJoinQuery(S1, S2,
           radius, windowSize, windowSlideStep, uGrid);
\end{lstlisting}

Usually, two types of indexes are used for spatial data: 1) Tree-based, 2) Grid-based. Unlike static data, stream tuples arrive and expire at a high velocity. Hence, tree-based spatial indexes are not suitable for it owing to their high maintenance cost \cite{Darius2009}. Therefore, to enable real-time processing of spatial data streams, a light weight logical grid index is introduced in this work. GeoFlink assigns grid-cell ID(s) to the incoming stream tuples based on which the objects are processed, pruned and/or distributed dynamically across the cluster nodes. GeoFlink currently supports the most commonly used spatial queries, i.e., spatial range, spatial $k$NN and spatial join on point data. It provides a user-friendly Java/Scala API to register spatial continuous queries (CQs). GeoFlink is an open source project and is available at Github\footnote{GeoFlink @ Github \url{https://github.com/aistairc/GeoFlink}}.

\begin{example}[Use case: Patients tracking] A city administration is interested in monitoring the movement of a number of their high-risk patients. Particularly, the administration is interested in knowing and notifying all the residents in real-time, if a patient happens to pass them within certain radius $r$. Let $S1$ and $S2$ denote the real-time ordinary residents' and patients' location stream, respectively, obtained through their smart-phones. Then, this query includes real-time join of $S1$ and $S2$, such that it outputs all the $p \in S1$ that lie within $r$ distance of any $q \in S2$. Code \ref{code:rangeQuery} shows the implementation of this real-time CQ using GeoFlink's spatial join. The details of each statement in the code is discussed in the following sections.
\end{example}


The main contributions of this work are summarized below:



\begin{itemize}
\item The core GeoFlink, which extends Apache Flink to support spatial data types, index and CQs. 
\item Grid-based spatial index for the efficient processing, pruning and distribution of spatial streams.
\item Grid-based spatial range, $k$NN and join queries.
\item An extensive experimental study on real spatial data streams. 
\end{itemize}

The rest of the paper is organized as follows: Sec. \ref{sec:relatedWork} presents related work. Sec. \ref{sec:flinkProgrammingModel} briefly discusses Apache Flink programming model. In Sec. \ref{sec:geoFlinkArchitecture}, GeoFlink architecture is presented. Secs. \ref{sec:spatialStreamLayer} and \ref{sec:spatialQueries} detail the Spatial Stream and the Spatial Query Processing layers of GeoFlink. In particular, Sec. \ref{sec:gridIndex} presents the GeoFlink's Gird index. In Sec. \ref{sec:experiments} detailed experimental study is presented while Sec. \ref{sec:conclusion} concludes our paper and highlights a few future directions.

\section{Related Work}
\label{sec:relatedWork}

Existing spatial data processing frameworks like ESRI ArcGIS \cite{ESRI}, PostGIS \cite{PostGIS} and QGIS \cite{QGIS2020} are built on relational DBMS and are therefore not scalable to handle huge data and throughput requirements. Besides, scalable spatial data processing frameworks, for instance, Hadoop GIS \cite{HadoopGIS2013}, Spatial Hadoop \cite{Spatialhadoop2015}, GeoSpark \cite{JiaYu2019}, Parallel Secondo \cite{Secondo2012} and GeoMesa \cite{GeoMesa2015}, cannot handle real-time processing of spatial data streams. Apache Spark \cite{SparkStreaming}, Apache Flink \cite{ApacheFlinkDoc} and similar distributed and horizontally scalable platforms support large-scale, real-time processing of data streams. However, they do not natively support spatial data processing and thus cannot process it efficiently. One can find a number of extensions of these platforms to support spatial data processing. GeoSpark \cite{JiaYu2019} processes spatial data by extending Spark’s native Resilient Distributed Dataset (RDD) to create Spatial RDD (SRDD) along with a Spatial Query Processing layer on top of the Spark API to run spatial queries on these SRDDs. For efficient spatial query processing, GeoSpark creates a local spatial index (Grid, R-tree) per RDD partition rather than a single global index. For re-usability, the created index can be cached on main memory and can also be persisted on secondary storage for later use. However, the index once created cannot be updated, and must be recreated to reflect any change in the dataset due to the immutable nature of RDDs. LocationSpark \cite{Tang2019LocationSparkID}, GeoMesa \cite{GeoMesa2015} and Spark GIS \cite{SparkGIS2017} are a few other spatial data processing frameworks developed on top of Apache Spark. All these frameworks, like the GeoSpark, do not support real-time stream processing as we do in GeoFlink.


For real-time queries, Apache Spark introduces Spark Streaming that relies on micro-batches to address latency concerns and mimic streaming computations. Latency is inversely proportional to batch size; however, the experimental evaluation in \cite{Karimov2018} shows that as the batch size is decreased to very small to mimic real-time streams, Apache Spark is prone to system crashes and exhibits lower throughput and fault tolerance. Furthermore, even with the micro-batching technique, Spark only approaches near real-time results at best, as data buffering latency still exists, however, miniscule. Other distributed streaming platforms worth considering are Apache Samza \cite{apacheSamza} and Apache Storm \cite{ApacheStorm}. Performance comparison by Fakrudeen et al. \cite{Fakrudeen2019} revealed that both the Samza and Storm demonstrate a lower throughput and reliability than Apache Flink \cite{ApacheFlinkDoc}. Thus, we extend Apache Flink, a distributed and scalable stream processing engine, to support real-time spatial stream processing. Furthermore, to enable efficient spatial query processing and data partitioning, a light-weight logical grid-based index is proposed.

\section{Flink Programming Model}
\label{sec:flinkProgrammingModel}

Apache Flink uses two data collections to represent data in a program: 1) DataSet: A static and bounded collection of tuples, 2) DataStream: A continuous and unbounded collection of tuples. However, both the collections are treated as streams internally. A Flink program consists of 3 building blocks: 1) Source, 2) Transformation(s), and 3) Sink. When executed, Flink programs are mapped to streaming dataflows, consisting of streams and transformation operators. Each dataflow starts with one or more sources and ends in one or more sinks. The dataflows resemble arbitrary directed acyclic graphs (DAGs); however, special forms of cycles are permitted via iteration constructs \cite{ApacheFlinkDoc}. By its very definition, dataflow processing offers low latency, thus for the real-time analytics use cases, Apache Flink is a natural choice.

Flink's DataStream API enables transformations like filter, map, reduce, keyby, aggregations, window, etc. on unbounded data streams and provides seamless connectivity with data sources and sinks like Apache Kafka (source/sink), Apache Cassandra (sink), etc. \cite{ApacheFlinkDoc}. Aggregates on streams (counts, sums, etc.), are scoped by windows, such as "count over the last 5 minutes", or "sum of the last 100 elements", since it is impossible to count all elements in a stream, because streams are in general unbounded. Windows can be time driven (e.g., every 30 seconds) or data driven (e.g., every 100 elements). One typically distinguishes different types of windows, such as tumbling windows (no overlap), sliding windows (with overlap), and session windows (punctuated by a gap of inactivity). When using windows, output is generated based on the complete window contents as it moves. While many operations in a dataflow simply look at one individual event at a time, some operations remember information across multiple events (for example window operators). These operations are called stateful.

Programs in Flink are inherently parallel and distributed. During execution, an operator is divided into one or more subtasks (operator instances) which are independent of one another and execute in different threads that may be on different machines or containers. The number of an operator's subtasks depends on the amount of its parallelism. A user can define the parallelism of each operator or set the maximum parallelism globally for all operators. Flink parallelism depends on the number of available task slots, where a good default number of task slots is equivalent to the number of CPU cores. In Flink, keys are responsible for the data distribution across the task slots or operator instances. All the tuples with the same key are guaranteed to be processed by a single operator instance. In addition, many of Flink's core data transformations like join, groupby, reduce and windowing require the data to be grouped on keys. Keying operations are enabled by \textit{KeyBy} operator, which logically partitions stream tuples with respect to their keys. Intelligent key assignment ensures the uniform data distribution among operator instances and hence leverage the performance offered by parallelism. 
 



Streams can transport data between two operators in a one-to-one (or forwarding) pattern, or in a redistributing pattern. One-to-one streams preserves partitioning and order of elements, while redistributing streams change the partitioning of streams. Each operator subtask sends data to different target subtasks, depending on the selected transformation. By default, each operator preserves the partitioning and order of the operator before it, thus preserving the source parallelism. While keying operations causes data reshuffling and distribution overhead, data forwarding may cause a load imbalance and even idling of cores that are not in use, thus not fully leveraging computation power of the entire cluster. Therefore, to guarantee efficient execution of queries, one must find the right balance between data redistribution and data forwarding. Furthermore, as parallel instances of operators cannot communicate with each other, data locality per instance must be ensured by the user.

\section{GeoFlink Architecture}
\label{sec:geoFlinkArchitecture}
\begin{figure}[!htb]
\centering
\includegraphics[width=0.3\textwidth]{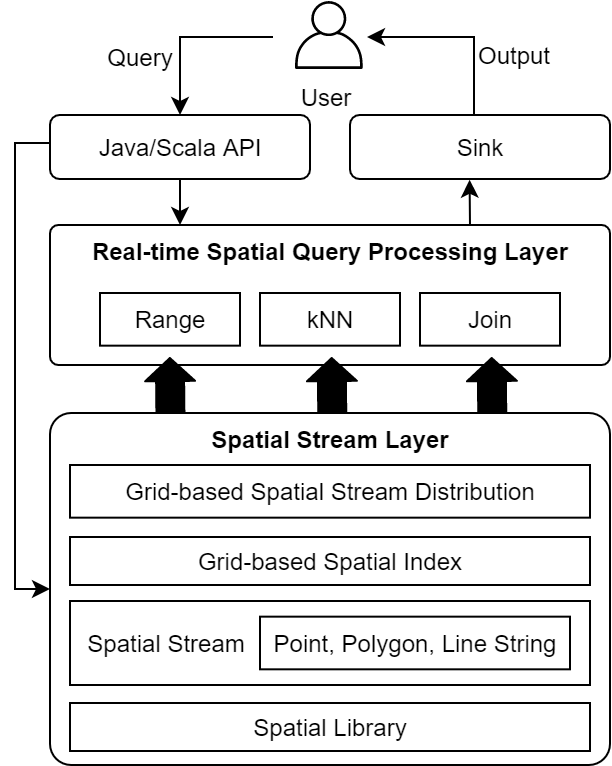}
\caption{GeoFlink architecture}
\label{fig:GeoFlinkArchitecture}
\end{figure}

Fig. \ref{fig:GeoFlinkArchitecture} shows the proposed GeoFlink architecture. Users can register queries to GeoFlink through a Java/Scala API and its output is available via a variety of sinks provided by Apache Flink. The GeoFlink architecture has two important layers: 1) Spatial Stream Layer and 2) Real-time Spatial Query Processing Layer.



\begin{flushleft}
\textbf{Spatial Stream Layer:} This layer is responsible for converting incoming data stream(s) into spatial data stream(s). Apache Flink treats spatial data stream as ordinary text stream, which may leads to its inefficient processing. GeoFlink converts it into spatial data stream of geometrical objects, i.e., point, line or polygon. Furthermore, this layer assigns Grid index keys to the spatial objects for their efficient distribution and processing.
\end{flushleft}

\begin{flushleft}
\textbf{Real-time Spatial Query Processing Layer:} This layer enables spatial queries' execution over spatial data streams. GeoFlink currently supports the most widely used spatial queries, i.e., spatial range, spatial $k$NN and spatial join queries over point objects. Users can use Java or Scala to write the spatial queries or custom applications. This layer makes extensive use of the Grid index for the efficient queries' execution. 
\end{flushleft}

\section{Spatial Stream Layer}
\label{sec:spatialStreamLayer}
This layer deals with the spatial stream construction and the Grid index (key) assignment to the stream tuples.

\subsection{Spatial Stream Indexing}
\label{subsec:spatialStreamIndexing}

\subsubsection{Tree vs. Grid Spatial Indexes}
The spatial data index structures can be classified into two broad categories: 1) Tree-based, and 2) Grid-based. Tree-based spatial indexes like R-tree, Quad-tree and KDB-tree can significantly speed-up the spatial query processing; however, their maintenance cost is high specially in the presence of heavy updates (insertions and deletions) \cite{RTrees2017}. On the other hand, grid-based indexes enable fast updates. However, they cannot answer queries as efficiently as tree-based indexes \cite{SpatialIndexing2009} \cite{Guting1994}. Since the GeoFlink is meant to support streaming applications with very high updates, the maintenance cost of the index employed has to be as small as possible. To this end, grid-based index seems to be a natural choice for GeoFlink.



\subsubsection{GeoFlink Grid Index}
\label{sec:gridIndex}

\begin{figure}[!htb]
\centering
\includegraphics[width=0.34\textwidth]{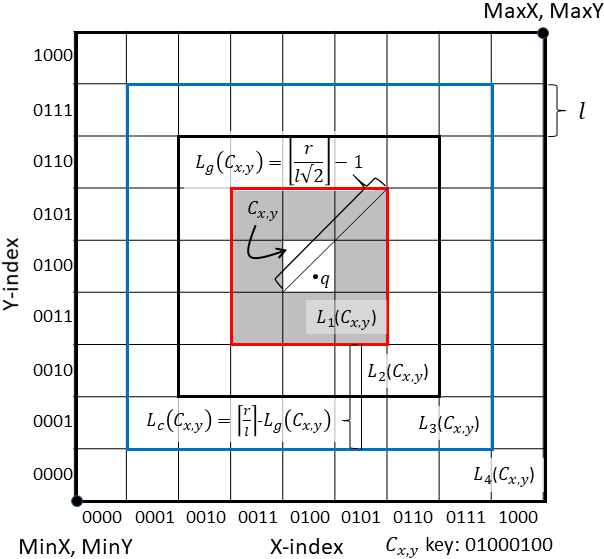}
\caption{GeoFlink grid index}
\label{fig:guaranteedCandidateLayers}
\end{figure}

A grid index \cite{Bentley1979} is a space-partitioned structure where a predefined area is divided into equal-sized cells of some fixed length $l$, as shown in Figure \ref{fig:guaranteedCandidateLayers}. 


The grid index used in this work is aimed at filtering/pruning objects during spatial queries' execution and helping the uniform distribution of spatial objects across GeoFlink's distributed cluster nodes. The Grid ($G$) is constructed by partitioning a 2D rectangular space, given by $(MinX, MinY), (MaxX, MaxY)$ $(MaxX-MinX = MaxY-MinY)$, into square shaped cells of length $l$. Here we assume that $G$'s boundary is known, which can be estimated through data stream's geographical location. Let $C_{x,y} \in G$ be a grid cell with indices $x$ and $y$, respectively, then $L_1(C_{x,y}), L_2(C_{x,y}),...,L_n(C_{x,y})$ denote its neighbouring layers, where $L_1(C_{x,y})$ is given by, $\{C_{u,v} | u = x \pm 1, v= y \pm 1, C_{u,v} \neq C_{x,y} \}$. Similarly, $L_2(C_{x,y}),...,L_n(C_{x,y})$ are defined. Each cell $C_{x,y} \in G$ is identified by its unique key obtained by concatenating its $x$ and $y$ indices. Figure \ref{fig:guaranteedCandidateLayers} shows a grid structure with a cell $C_{x,y}$, its unique key, and its layers $L_1(C_{x,y}), L_2(C_{x,y}),...,L_4(C_{x,y})$.

Within GeoFlink, each stream tuple is assigned key(s) on its arrival, depending upon the $G$ cell(s) it belongs. A geometrical object belongs to a cell $c$ if its coordinates lie within the boundary of $c$. In this work, we assume that a \textit{point} can belongs to only one cell, whereas, a \textit{line} and \textit{polygon} can belong to multiple cells depending upon their sizes and positions. Hence, a single key is assigned to a point whereas an array of key(s) may need to be assigned to a line and polygon. Since the focus of this work is point object, one key is assigned per stream tuple. Let $S$ denotes a spatial stream, then the coordinates of a tuple $s \in S$ are given by $s.x$ and $s.y$. Given the grid boundary $(MinX, MinY), (MaxX, MaxY)$ and grid size $m$, the cell length is computed as $l = \frac{MaxX - MinX}{m}$, and the key of a $s \in S$ is obtained as $xIndex = \lfloor \frac{s.x - G.MinX}{l} \rfloor$, $yIndex = \lfloor \frac{ s.y - G.MinY }{l} \rfloor$, and $s.key = xIndex \odot  yIndex$. Where $xIndex$ and $yIndex$ are fixed length indices of bit length $n$ and $\odot$ denotes a concatenation operator. For instance, let $G$ is given by $(MinX, MinY)=(0,0)$, $(MaxX, MaxY)=(90,90)$ and $m=9$, then, for a $s \in S$ with coordinates $(25, 42)$ and $n=4$, $s.key$ is given by: $xIndex = 0010$, $yIndex = 0100$, $s.key = 00100100$.

The grid-based index used in this work is logical, that is, it only assigns a key to the incoming streaming tuples or moving objects. Besides, no physical data structure is needed, hence no update is required when a stream tuple expires or an updated object location is received. This makes our grid index fast and memory efficient. In GeoFlink, a grid index is constructed through \textit{UniformGrid} class.

\begin{lstlisting}[mathescape=true]
UniformGrid $G$ = new UniformGrid
                     (GridSize, MinX, MaxX, MinY, MaxY);
\end{lstlisting}

\begin{flushleft}
where \textit{GridSize=50} generates a grid of 50x50 cells, with the bottom-left (MinX, MinY) and top-right (MaxX, MaxY) coordinates, respectively.
\end{flushleft}

\subsection{Spatial Objects Support}
\label{subsec:spatialObjectsSupport}
GeoFlink currently supports \textit{GeoJSON} and \textit{CSV} input stream formats from Apache Kafka and \textit{Point} type spatial objects. However, we are working on its extension to support other input formats and spatial objects including lines and polygons.

GeoFlink user needs to make an appropriate Apache Kafka connection by specifying the kafka \textit{topic name} and bootstrap server(s). Once the connection is established, the user can construct spatial stream from GeoJSON input stream by utilizing the \textit{PointStream} method of the GeoFlink's \textit{SpatialStream} class.

\begin{lstlisting}[mathescape=true]
DataStream<Point> $S$ = SpatialStream.
             PointStream(geoJSONStream, "GeoJSON", $G$);
\end{lstlisting}

%

\subsection{Spatial Stream Partitioning}
Uniform partitioning of data across distributed cluster nodes plays a vital role in efficient query processing. As discussed in Section \ref{sec:flinkProgrammingModel}, Apache Flink \textit{keyBy} transformation logically partitions a stream into disjoint partitions in such a way that all the tuples with the same key are assigned to the same partition or to the same operator instance. If the number of unique keys are larger than the amount of parallelism, multiple keys are assigned to a single operator instance.

To enable uniform data partitioning in GeoFlink, which takes into account data spatial proximity, grid index is used. As discussed earlier, GeoFlink assigns a grid  cell \textit{key} to each incoming stream tuple based on its spatial location. Since all the spatially close tuples belong to a single grid cell, thus, are assigned the same key, which is used by the Flink's \textit{keyBy} operator for stream distribution. It is good to have the number of keys greater than or equal to the amount of parallelism, to enable the Flink to distribute data uniformly.

It is worth mentioning that, GeoFlink receives distributed data streams from distributed messaging system, for instance, Apache Kafka \cite{KafkaStreaming}. To enable uniform distribution of incoming data stream across GeoFlink cluster nodes, right configuration is needed. Many times, improper configuration becomes a serious bottleneck, resulting in reduced system throughput. For instance, assuming that Kafka is used as a data source then its topic must be partitioned keeping in view the Flink cluster parallelism, i.e., the number of topic partitions must be greater than or equal to the Flink parallelism so that no GeoFlink operators' instance remain idle while fetching the data. The detailed discussion on the configuration is outside the scope of this work.


\section{Spatial Query Processing Layer}
\label{sec:spatialQueries}
This layer provides support for all the basic spatial operators required by most of the spatial data processing and analysis applications. The supported queries include spatial range, spatial $k$NN and spatial join queries. All the queries discussed in this section are window-based and are continuous in nature, i.e., they generate window-based continuous results on continuous data stream. Namely, one output is generated per window aggregation as it slides. Due to the stateless nature of most of the Flink's transformations, the queries' results are computed in a non-incremental fashion, i.e., the results are generated using all the objects in each window without considering the past window results. To reduce the query execution cost, GeoFlink makes use of the grid index. Unless stated otherwise, in the following, the notations $S$, $q$, $r$, and $l$ are used for spatial data stream, query object, query radius and grid cell length, respectively. Furthermore, window size and window slide step (also known as window parameters) are denoted by $W_n$ and $W_s$, respectively. Since most of the spatial queries deal with neighbourhood computation, we define $r$-neighbors of $q$ as follows.

\begin{definition} [$r$-neighbors($q$)]
Geometrical objects that lie within the radius $r$ of $q$.
\end{definition}

One traditional and a very effective approach to reduce the computation cost of a query is to prune out the objects which cannot be an $r$-neighbor($q$). Given a cell $C_{x,y} \in G$ containing $q$ as shown in Figure \ref{fig:guaranteedCandidateLayers}, the pruning cell layers are defined as follows:

\begin{itemize}
\item \textbf{Guaranteed Layers ($L_g(C_{x,y})$):} The objects in this layer are guaranteed to be an $r$-neighbor($q$). 

$L_g(C_{x,y}) = \{C_{u,v} | u = x \pm g, v = y \pm g, C_{u,v} \neq C_{x,y} \}$, where $g = \lfloor \frac{r}{l \sqrt{2}} \rfloor -1$.

\item \textbf{Candidate Layers ($L_c(C_{x,y})$):} The objects in this layer may or may not be an $r$-neighbor($q$). Hence, require (distance) evaluation. 

$L_c(C_{x,y}) = \{C_{u,v} | u = x \pm c, v = y \pm c, C_{u,v} \notin L_g(C_{x,y}), C_{u,v} \neq C_{x,y} \}$, where $c = \lceil \frac{r}{l} \rceil$.

\item \textbf{Non-neighbouring Layers ($L_n(C_{x,y}):$ others):} The objects in this layer cannot be an $r$-neighbor($q$). Hence, can be safely pruned.
\end{itemize}

%
%


The cells in the layers $L_g(C_{x,y})$, $L_c(C_{x,y})$ and $L_n(C_{x,y})$ are disjoint. In the following, we call the objects belonging to the $L_g(C_{x,y})$, $L_c(C_{x,y})$ and $L_n(C_{x,y})$ layers as the guaranteed-, candidate-, and non-neighbors of $q$, respectively.

\begin{example}
Let the grid ($G$) of Fig. \ref{fig:guaranteedCandidateLayers} is given by $(MinX, MinY)$ $=(0,0)$, $(MaxX, MaxY)=$ $(90,90)$, then $l=10$. Assuming that $q$ lies in the cell $C_{x,y}$ and let $r=30$. Then, $g = \lfloor \frac{r}{l \sqrt{2}} \rfloor -1 = 1$ and the guaranteed layer is given by the layers within red boundary in Fig. \ref{fig:guaranteedCandidateLayers}, excluding the cell $C_{x,y}$. All the objects in this layer are guaranteed-neighbors of $q$ results. Similarly, $c = \lceil \frac{r}{l} \rceil = 3$ and the candidate layer is given by the layers within blue boundary in the figure, excluding the guaranteed layer and $C_{x,y}$. All the objects in this layer are candidate-neighbors of $q$ and must be evaluated using distance function to find if they are $r$-neighbors($q$). Rest of the layers contain only non-neighbors of $q$. 
\end{example}

%
%
%

%

\subsection{Spatial Range Query}
\label{subsec:rangeQuery}


\begin{definition}[Spatial Range Query]
Given $S$, $q$, $r$, $W_n$ and $W_s$, range query returns the $r$-neighbors($q$) in $S$ for each aggregation window.
\end{definition}

A spatial range query returns all the $s \in S$ in a window, that lie within the $r$-distance of $q$. The query results are generated periodically based on $W_n$ and $W_s$. Such a query can be easily distributed and parallelized, i.e., the $S$ tuples can be divided across distributed cluster nodes, where each tuple is checked for $r$-neighbors($q$). This is a naive approach and require distance computation between all $s \in S$ and $q$, which can be computationally expensive, specially when the distance function is expensive, for instance, road distance.

\begin{figure}[!htb]
\centering
\includegraphics[width=0.4\textwidth]{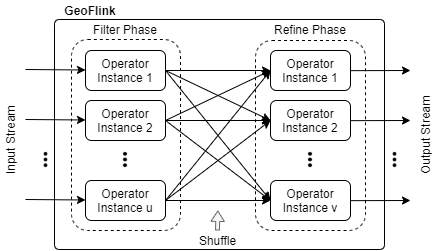}
\caption{Spatial Range Query Data Flow}
\label{fig:rangeQuery}
\end{figure}

A more efficient way is to prune out the objects which cannot be part of the query result, thus reducing the number of distance computations and the query processing cost. An effective pruning requires some index structure to identify the objects which can be safely pruned. Hence, we propose a grid-based spatial range query consisting of \textit{Filter} and \textit{Refine} phases as shown in Figure \ref{fig:rangeQuery}. Herein the \textit{Filter} phase prunes out the objects which cannot be part of the query output and the \textit{Refine} phase evaluates the un-pruned objects using distance function. Precisely, given $q$ and $r$, each GeoFlink node computes $L_g(C_q)$ and $L_c(C_q)$ sets, where $C_q$ denotes the cell containing $q$. The \textit{Filter} phase prunes out the $S$ tuples which are not part of $L_g(C_q)$ or $L_c(C_q)$. Filtered stream is then shuffled to keep the data balanced across the nodes in the \textit{Refine} phase. Since the $S$ objects corresponding to $L_g(C_q)$ are guaranteed $r$-neighbour($q$), only the objects corresponding to $L_c(C_q)$ are checked for $r$-neighbour($q$) using distance function in the \textit{Refine} phase. From Fig. \ref{fig:rangeQuery}, the number of operator instances in filter and refine phases are $u$ and $v$, respectively, where $u \geq v$. To execute a spatial range query via GeoFlink, \textit{SpatialRangeQuery} method of \textit{RangeQuery} class is used.

%
%
%
%
%
%
%
%

%
%

\begin{lstlisting}[mathescape=true]
DataStream<Point> rangeOut = RangeQuery.
              SpatialRangeQuery($S$, $q$, $r$, $W_n$, $W_s$, $G$); 
\end{lstlisting}


\subsection{Spatial $k$NN Query}
\label{subsec:kNNQuery}
\begin{definition}[Spatial $k$NN Query]
Given $S$, $q$, $r$, $W_n$, $W_s$ and a positive integer $k$, $k$NN query returns the nearest $k$ $r$-neighbors($q$) in $S$ for each aggregation window. If less than $k$ neighbors exists then all the $r$-neighbors($q$) are returned. 
\end{definition}


To find $k$NN naively, distances between all $s \in S$ in a window and $q$ are computed and the $k$ nearest objects to $q$ are returned for each window. This query can be easily distributed and parallelized, i.e., the $S$ tuples can be divided across the cluster nodes, where each node computes and maintains its $k$ nearest neighbors. The $k$NNs are then merged and sorted on a single cluster node to generate the true $k$NNs per window. However, this approach is expensive due to the large number of distance computations.



\begin{figure}[!htb]
\centering
\includegraphics[width=0.48\textwidth]{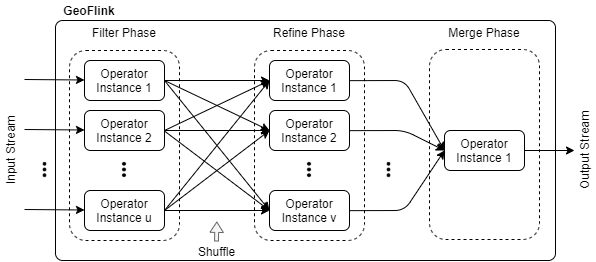}
\caption{$k$NN Query Data Flow}
\label{fig:kNNQuery}
\end{figure}

\begin{figure*}[!htb]
\centering
\includegraphics[width=0.75\textwidth]{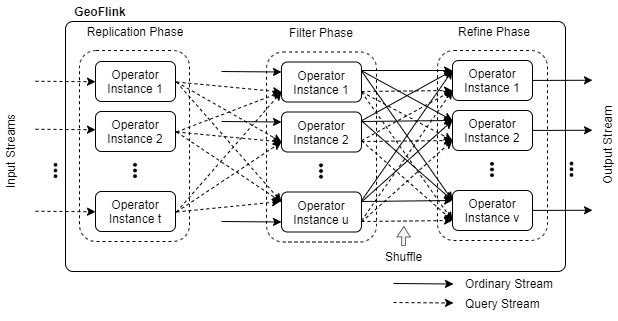}
\caption{Spatial Join Query Data Flow}
\label{fig:spatialJoin}
\end{figure*}

This work presents an efficient grid-based $k$NN approach, consisting of \textit{Filter}, \textit{Refine} and \textit{Merge} phases as shown in Figure \ref{fig:kNNQuery}. In the \textit{Filter} phase, the objects in the non-neighbouring layers are pruned. The \textit{Refine} phase evaluates the objects in the guaranteed and candidate layers using distance function. The \textit{Merge} phase is responsible for integrating the $k$NNs from distributed cluster nodes and sorting them to obtain true $k$NNs. Precisely, given $q$ and $r$, GeoFlink nodes compute $L_g(C_q)$ and $L_c(C_q)$ sets, where $C_q$ denotes the cell containing $q$. The \textit{Filter} phase prunes out the $S$ tuples which are not part of $L_g(C_q)$ or $L_c(C_q)$. Filtered stream is then shuffled to keep the data balanced across the nodes in the \textit{Refine} phase. To compute the $k$NNs in the \textit{Refine} phase, distances of the nearest $k$ $r$-neighbors($q$) are maintained on a heap. The heap's root points to the $k^{th}$ nearest object and is updated as a new candidate $k$NN is found. The \textit{Refine} phase is executed in a distributed fashion, i.e., each node computes its own copy of $k$NNs. The \textit{Merge} phase receives $k$NNs from all the distributed nodes for each window, integrates and sorts them to obtain true $k$NNs. To execute a spatial $k$NN query in GeoFlink, \textit{SpatialKNNQuery} method of the \textit{KNNQuery} class is used.

%

\begin{lstlisting}[mathescape=true]
DataStream <PriorityQueue<Tuple2<Point, Double>>> 
kNNOut = KNNQuery.SpatialKNNQuery($S$, $q$, $r$, $k$, $W_n$, $W_s$, $G$);
\end{lstlisting}

Please note that the output of the $k$NN query is a stream of sorted lists with respect to the distance from $q$, where each list consists of $k$NNs corresponding to a window.

\subsection{Spatial Join Query}

\begin{definition}\label{def:spatialJoin}(Spatial Join Query)
Given $r$, $W_n$, $W_s$, and two streams $S1$ (Ordinary stream) and $S2$ (Query stream), spatial join query returns all the $r$-neighbors($q_i$) in $S1$ for each aggregation window, where $q_i \in S2$.
\end{definition}

\begin{figure*}[!htb]
\centering 
\rotatebox{90}{\scriptsize Higher better $\rightarrow$}
\subfigure[Part 1][Varying grid size]{\includegraphics[width=0.23\textwidth]{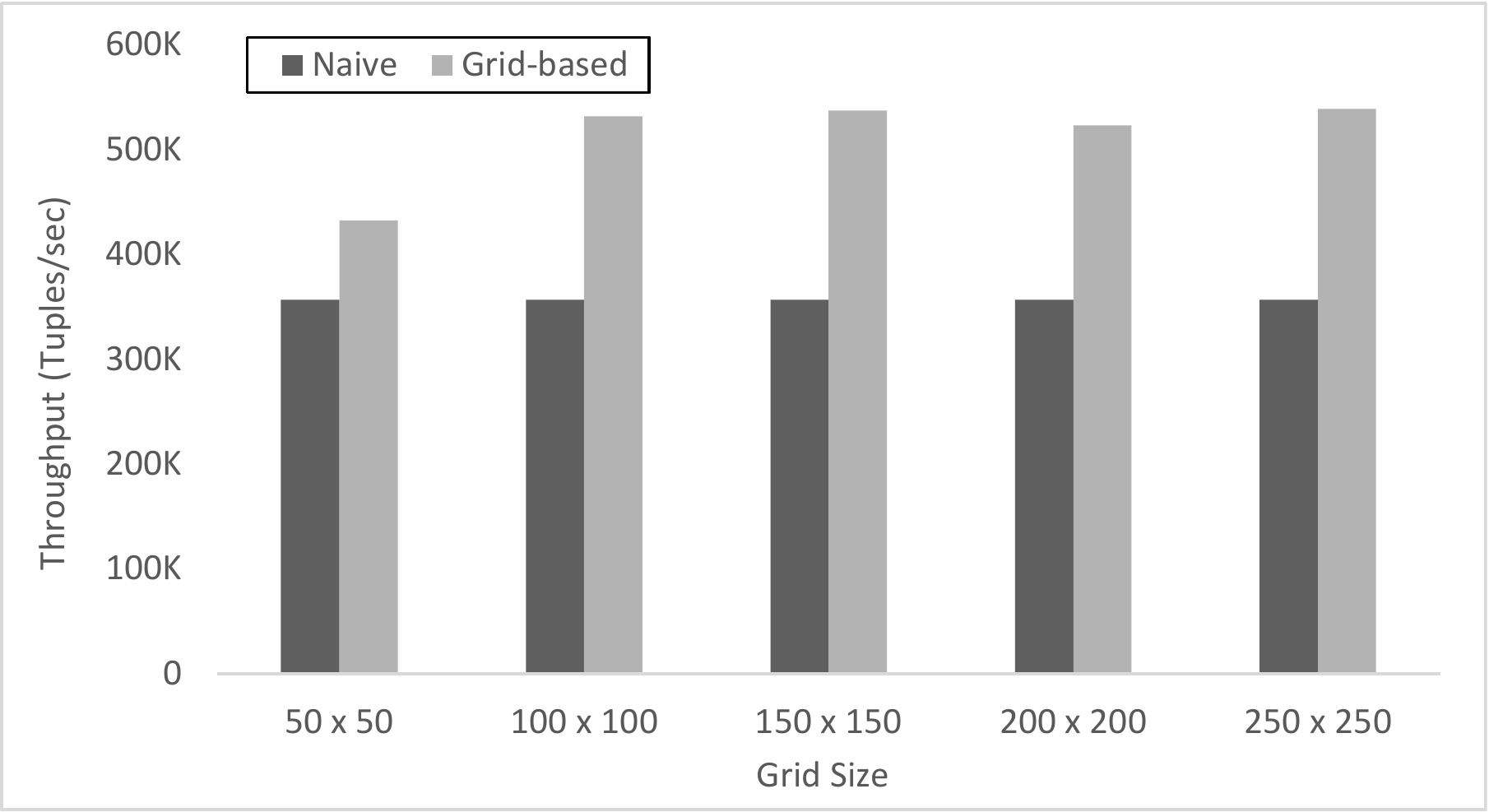} \label{fig:rangeGridSize}}
\subfigure[Part 2][Varying query radius]{\includegraphics[width=0.23\textwidth]{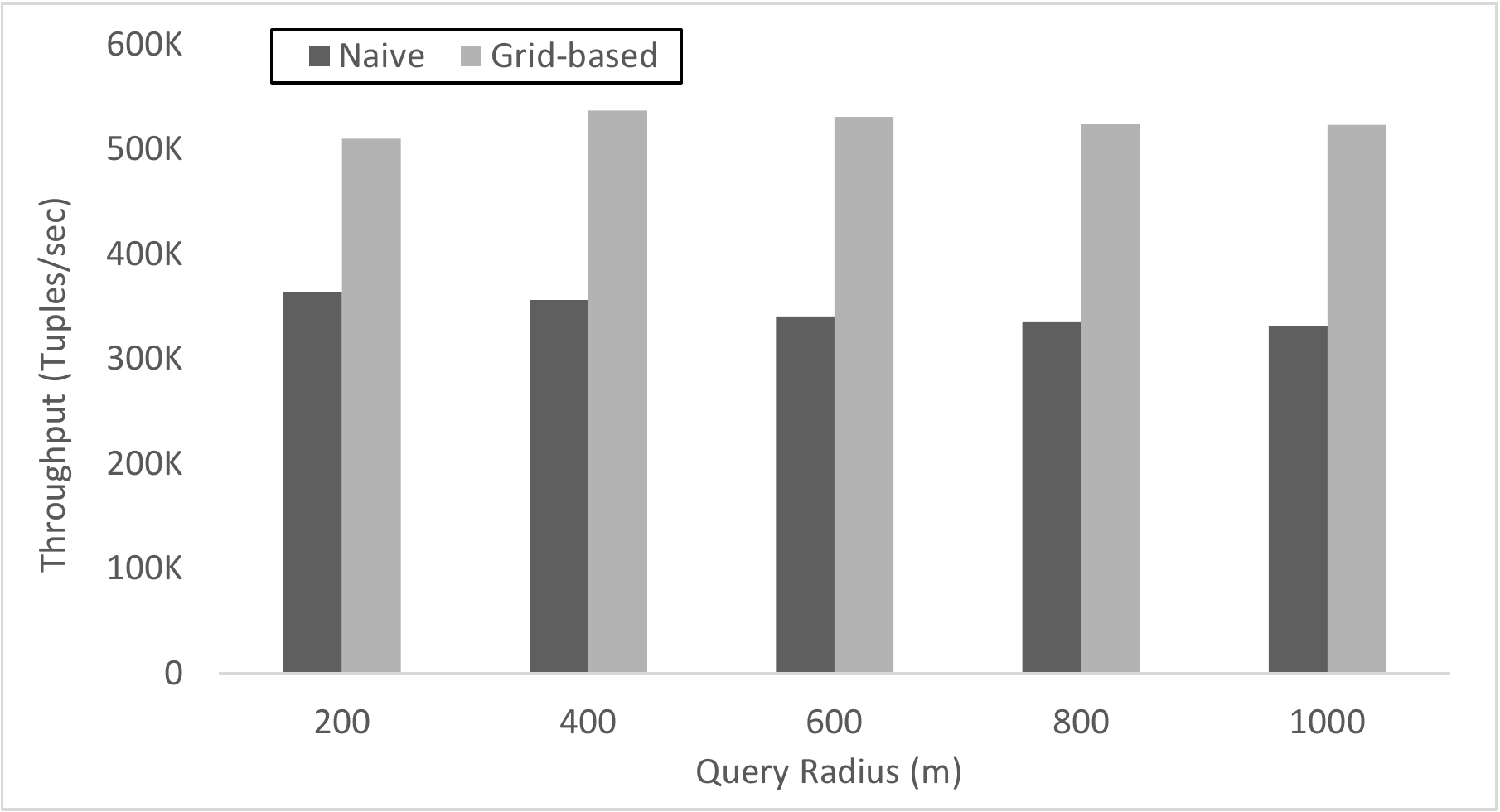} \label{fig:rangeQueryRadius}}
\subfigure[Part 3][Varying window size]{\includegraphics[width=0.23\textwidth]{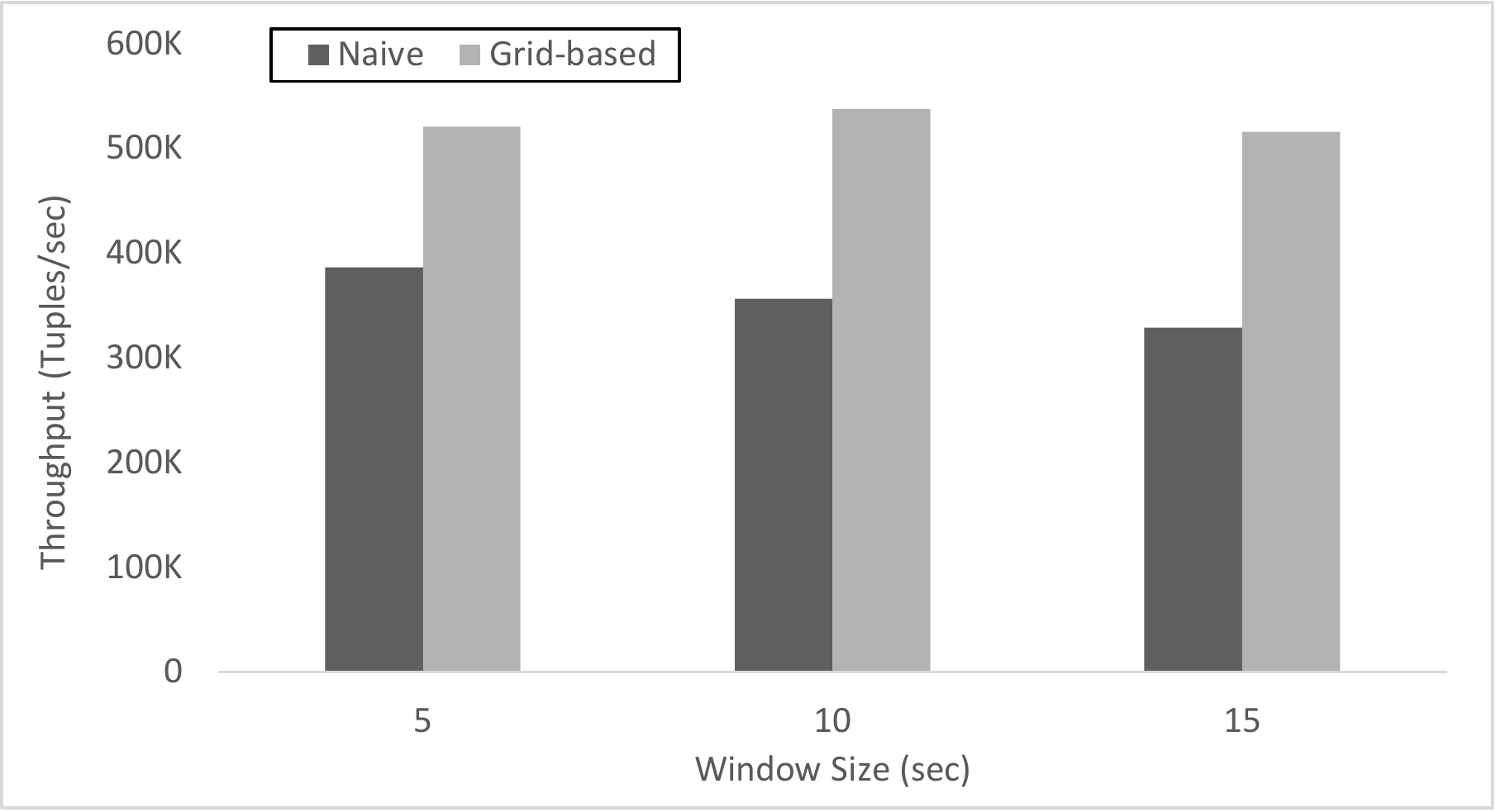} \label{fig:rangeWinSize}}
\subfigure[Part 4][Varying window slide step]{\includegraphics[width=0.23\textwidth]{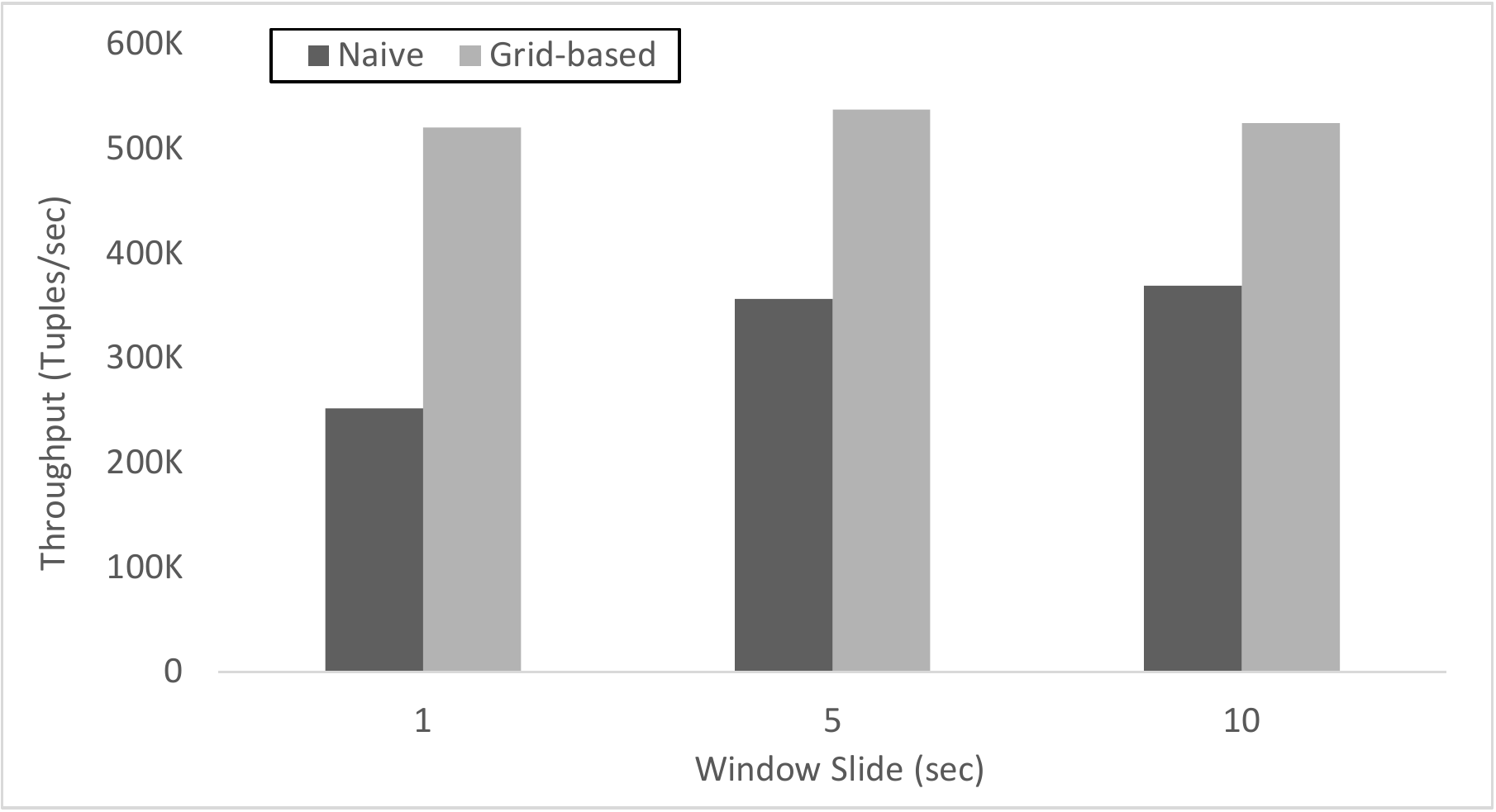} \label{fig:rangeWinSlide}}
\caption{Spatial range query} 
\label{fig:graphsRangeQuery}
\end{figure*}

\begin{figure*}[!htb]
\centering 
\rotatebox{90}{\scriptsize Higher better $\rightarrow$}
\subfigure[Part 1][Varying grid size]{\includegraphics[width=0.23\textwidth]{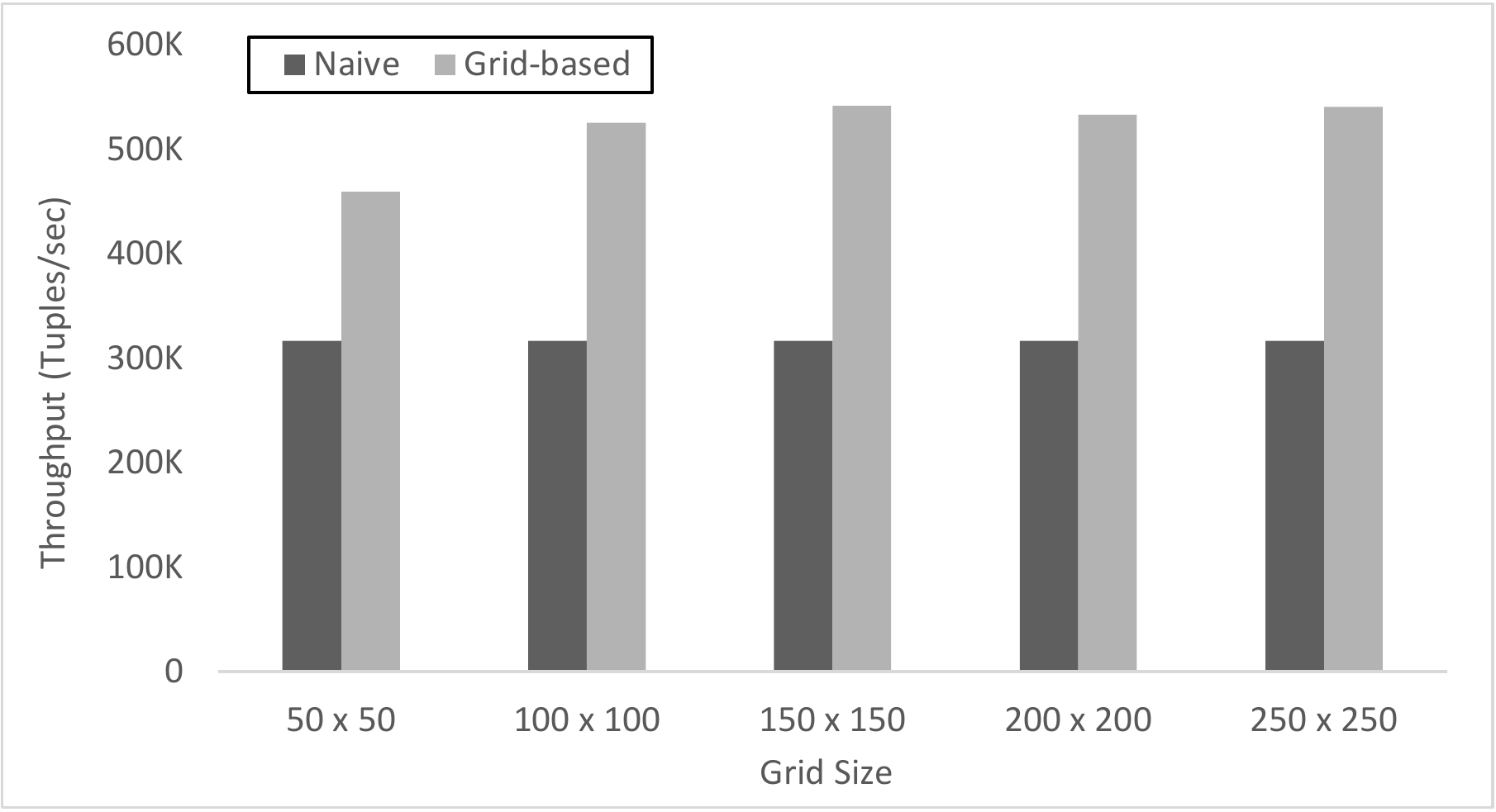} \label{fig:kNNGridSize}}
\subfigure[Part 2][Varying $k$]{\includegraphics[width=0.23\textwidth]{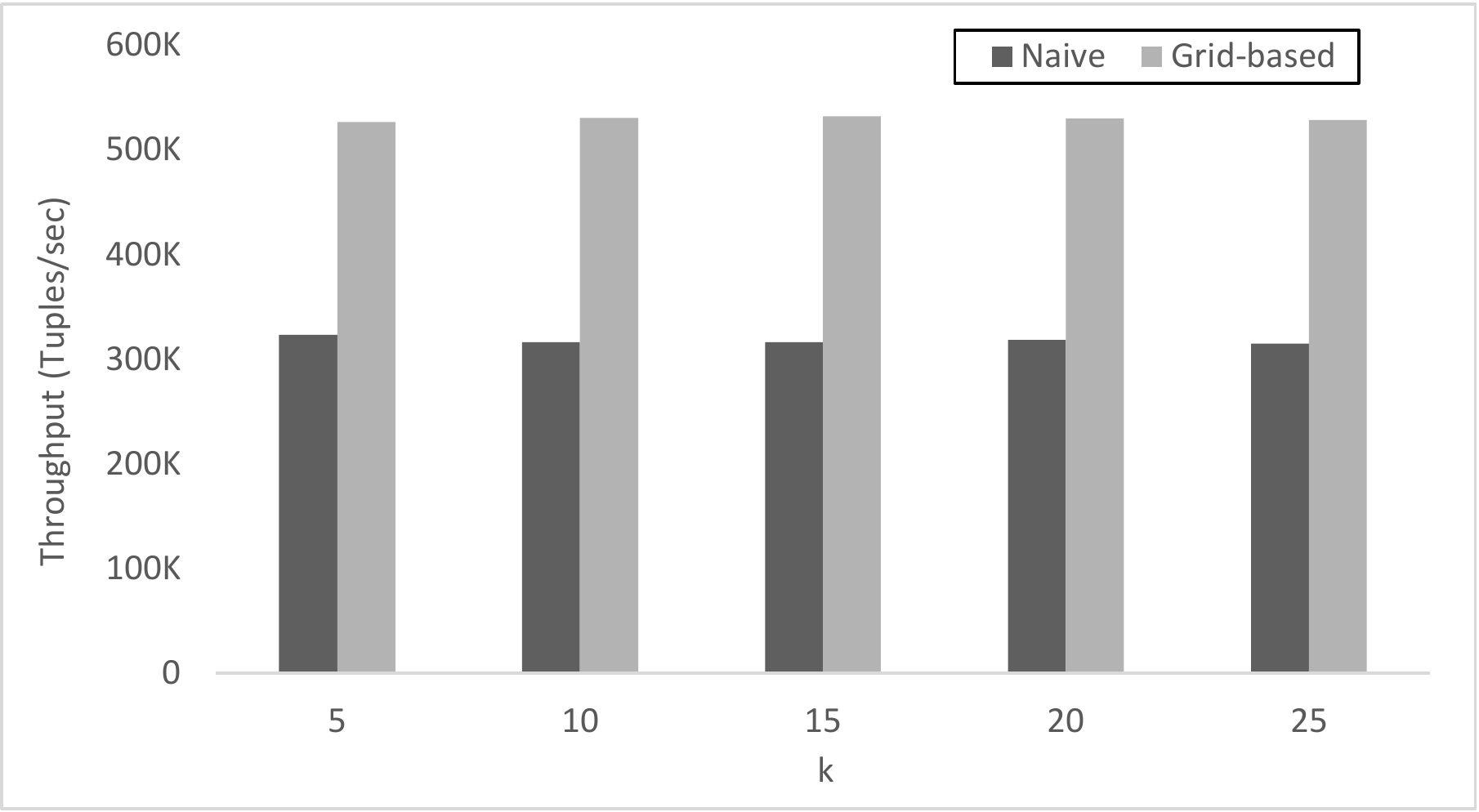} \label{fig:kNNk}}
\subfigure[Part 3][Varying window size]{\includegraphics[width=0.23\textwidth]{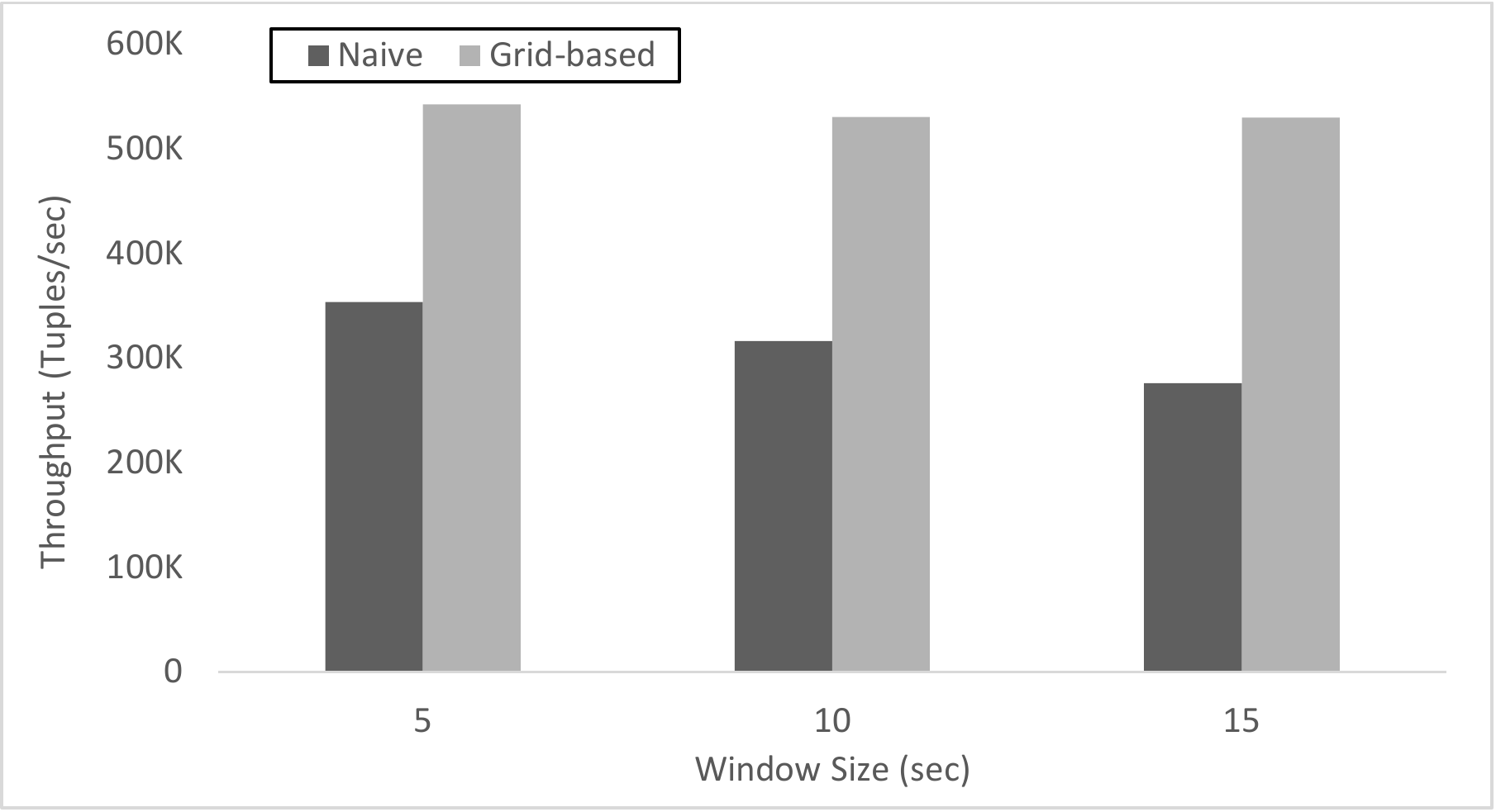} \label{fig:kNNWinSize}}
\subfigure[Part 4][Varying window slide step]{\includegraphics[width=0.23\textwidth]{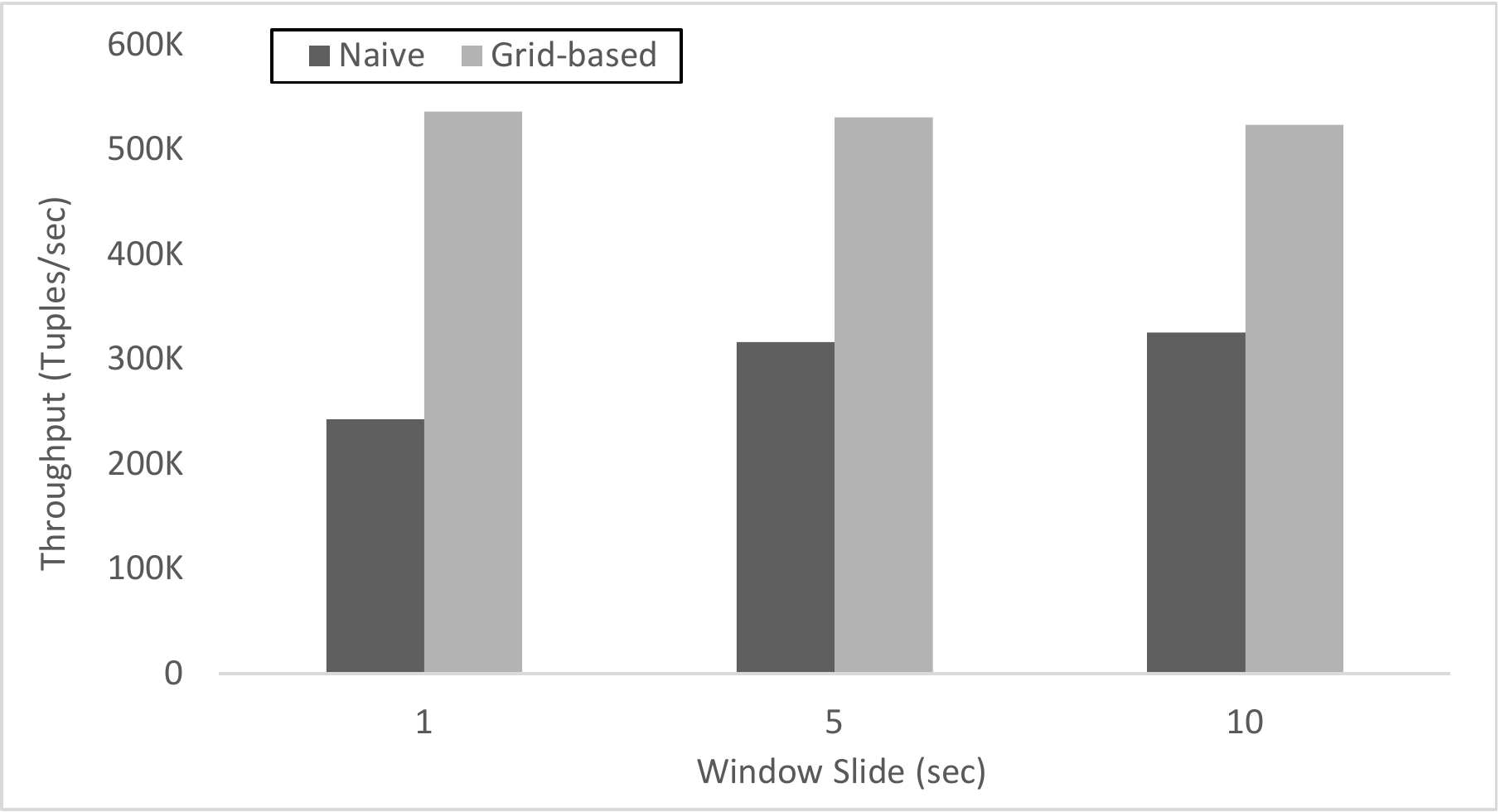} \label{fig:kNNWinSlide}}
\caption{Spatial kNN query} 
\label{fig:graphskNNQuery}
\end{figure*}

\begin{figure*}[!htb]
\centering 
\rotatebox{90}{\scriptsize Higher better $\rightarrow$}
\subfigure[Part 1][Varying grid size]{\includegraphics[width=0.23\textwidth]{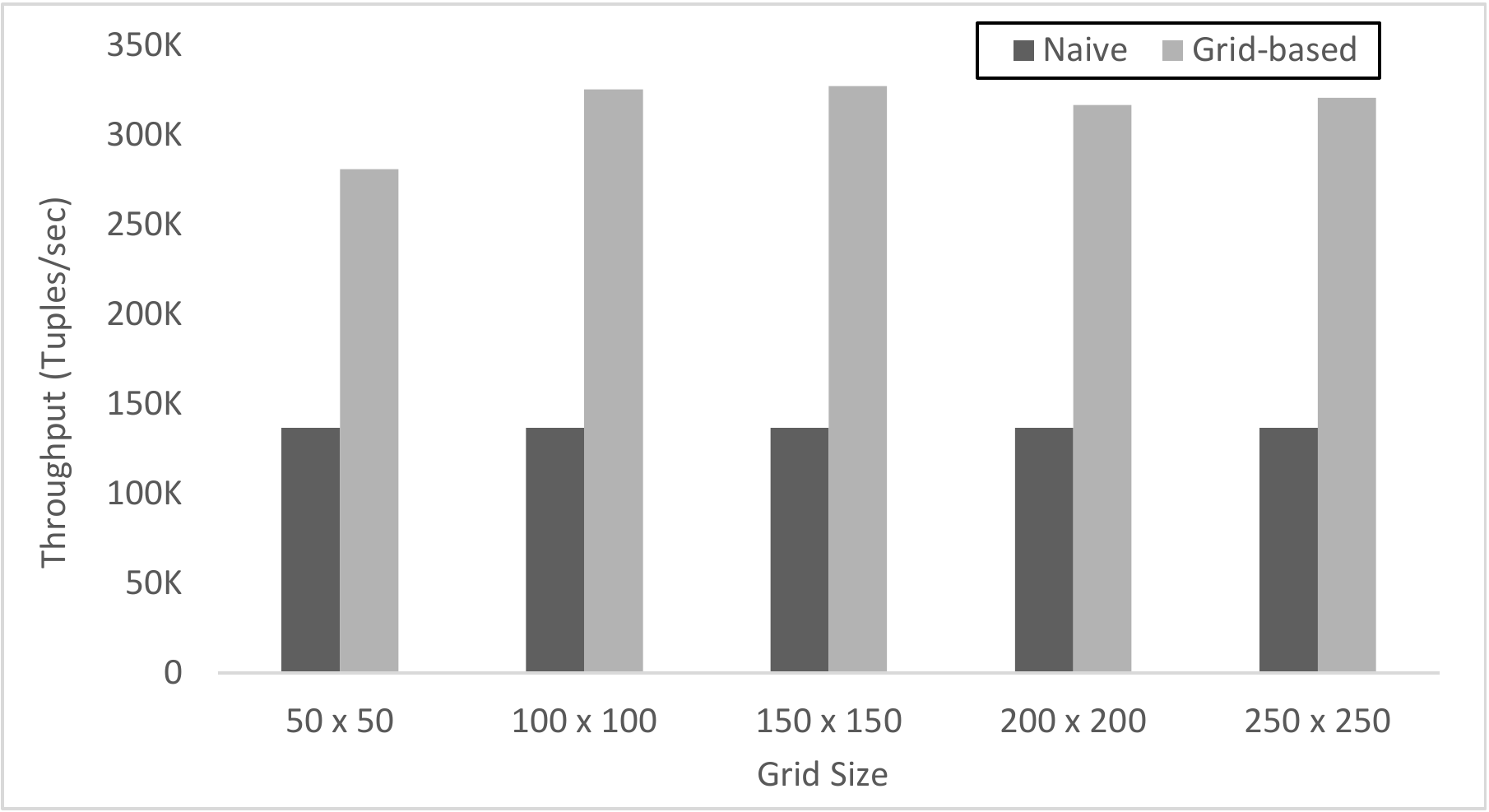} \label{fig:joinGridSize}}
\subfigure[Part 2][Varying query stream arrival rate]{\includegraphics[width=0.23\textwidth]{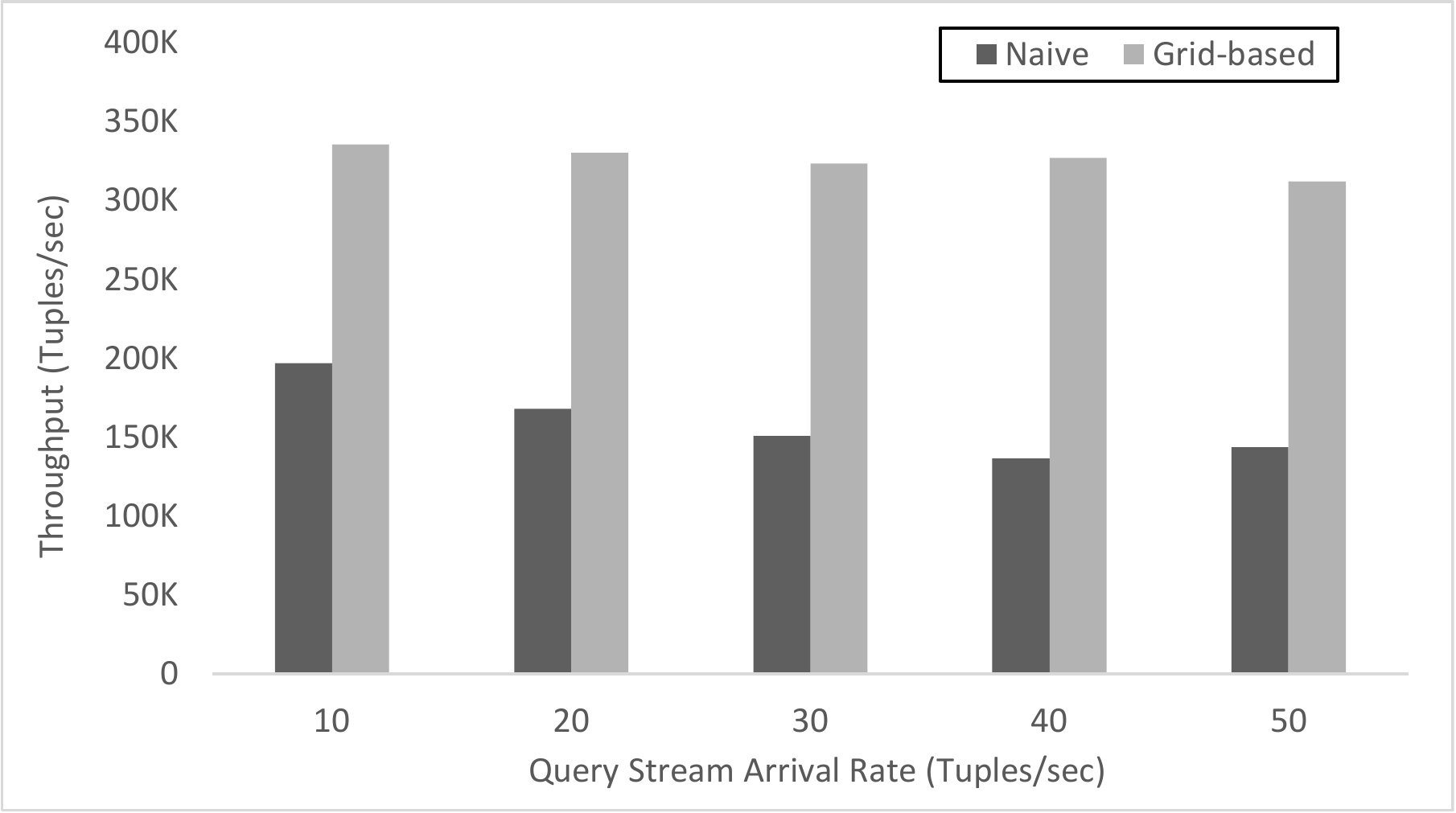} \label{fig:joinQueryStream}}
\subfigure[Part 3][Varying window size]{\includegraphics[width=0.23\textwidth]{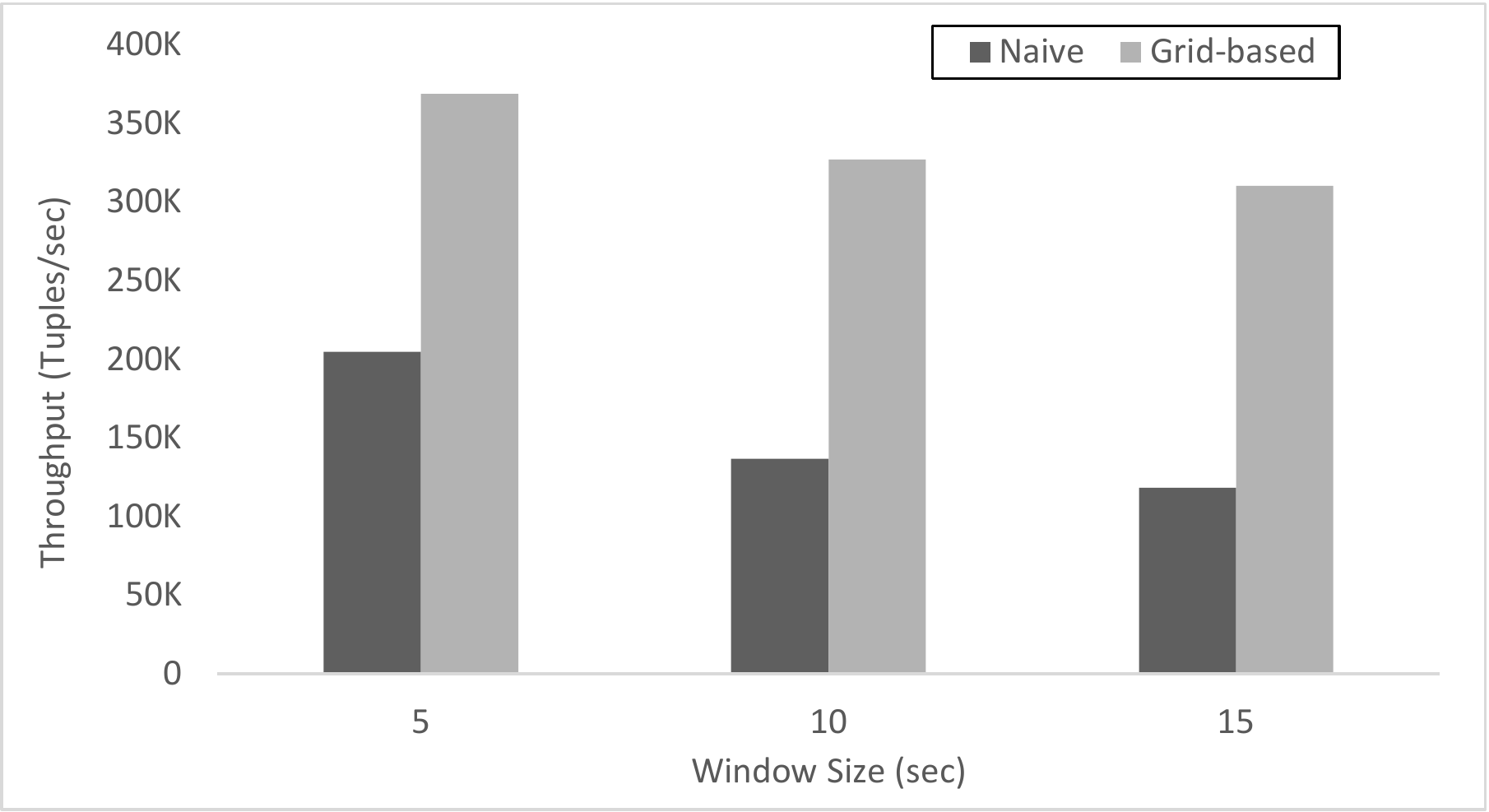} \label{fig:joinWinSize}}
\subfigure[Part 4][Varying window slide step]{\includegraphics[width=0.23\textwidth]{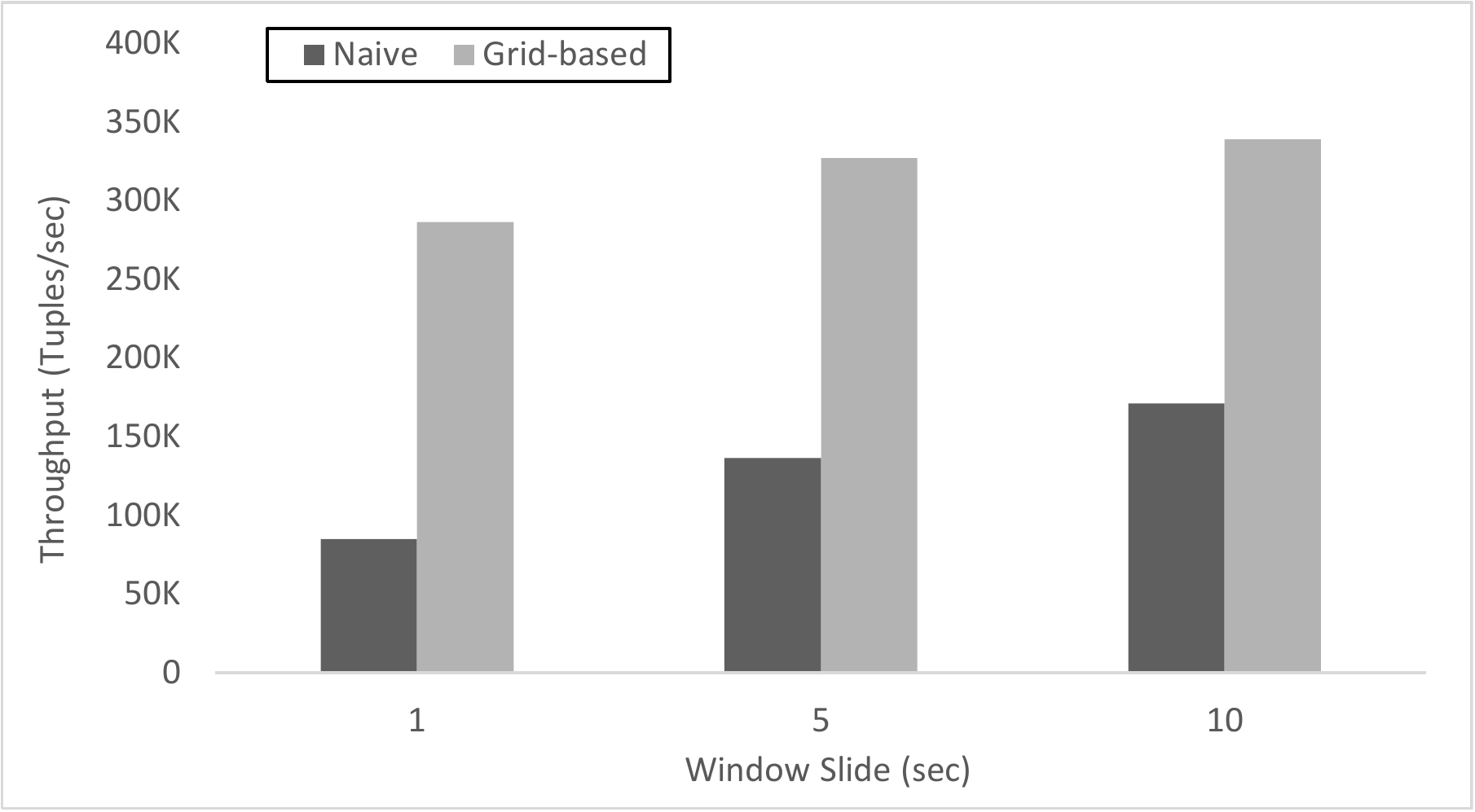} \label{fig:joinWinSlide}}
\caption{Spatial join query} 
\label{fig:graphsJoinQuery}
\end{figure*}

Spatial join is an expensive operation, where each tuple of query stream must be checked against every tuple of ordinary stream. To achieve this using a naive approach, low rate stream is replicated on all the cluster nodes whereas high rate stream is divided across them. However, this involves a large number of distance computations equivalent to the Cartesian product of the two streams and heavy shuffling of the tuples.

Hence, we propose an efficient grid index based spatial join. Figure \ref{fig:spatialJoin} gives an overview of the GeoFlink spatial join. The proposed spatial join consists of the following three phases: 1) Replication phase, 2) Filter phase, and 3) Refine phase. Let $S1$ and $S2$ denote an ordinary and a query stream, respectively. Assuming that $C_q$ denotes a cell containing a query object $q$, then given $r$, the \textit{Replication} phase computes the $L_g(C_{q})$ and $L_c(C_{q})$ layers for each $q \in S2$ in the current window. Next, the $q \in S2$ are replicated in such a way that each replicated point is assigned keys from the sets $L_g(C_{q})$ and $L_c(C_{q})$. We denote the replicated query stream by $S2'$. Next, we make use of Apache Flink's key-based join transformation to join the two streams, i.e., $S1$ and $S2'$. The Flink's key-based join enables the tuples from the two streams with the same key to land on the same operator instance. This causes the join to be evaluated only between $q \in S2'$ and $p \in S1$ belonging to the cells in $L_g(C_{q})$ and $L_c(C_{q})$, while filtering out the non-neighbors of $q$. In Figure \ref{fig:spatialJoin}, this corresponds to the \textit{Filter} phase. In the \textit{Refine} phase, since the $p \in S1$ corresponding to $L_g(C_{q})$ are guaranteed to be part of the join output, they are sent to the output directly without distance evaluation. However, for $p \in S1$ corresponding to $L_c(C_{q})$, distance-based evaluation is done to find if $p \in S1$ is an $r$-neighbors($q$), where $q \in S2'$. To execute a spatial join query via GeoFlink, \textit{SpatialJoinQuery} method of the \textit{JoinQuery} class is used.

\begin{lstlisting}[mathescape=true]
DataStream<Tuple2<String,String>> joinOut =         
    JoinQuery.SpatialJoinQuery($S$, $q$, $r$, $W_n$, $W_s$, $G$);
\end{lstlisting}

%

\begin{example}
Let $S1$ and $S2$ denote ordinary and query streams, respectively. We would like to perform the spatial window-join between these streams. Assuming that the window contains twenty $S1$ points $p1, p2, ..., p20$ and two $S2$ points $q1, q2$. Let $S1$ points are assigned cell-IDs (keys) based on their coordinates as follows: $c1 -> p1, p2, p3$, $c2 -> p4, p5, p6, p7$, $c3 -> p8, p9$, $c4 -> p10, p11, p12$, $c5 -> p13, p14, p15$ and $c6 -> p16, p17, p18, p19, p20$. Assuming that $q1$ and $q2$ belong to cells $C_{q1}$ and $C_{q2}$, respectively, and their neighbouring cells in candidate layers are given by $L_c(C_{q1}) = \{c2, c3\}$ and $L_c(C_{q2}) = \{c3, c5\}$, respectively. For the sake of simplicity in this example, we assume that the guaranteed layer does not exist. To enable our grid-based spatial join, $S2$ objects are replicated and assigned cell-IDs as: $c2 -> q1$, $c3 -> q1, q2$ and $c5 -> q2$. Let $S2'$ denotes the replicated query stream, then the spatial join between $S1$ and $S2'$ is executed in GeoFlink using three join operator instances handling keys $c2$, $c3$ and $c5$, respectively, as: \textit{Join Instance 1)} $q1$ join  $p4, p5, p6, p7$, \textit{Join Instance 2)} $q1, q2$ join  $p8, p9$ and  \textit{Join Instance 3)} $q2$ join  $p13, p14, p15$. Since the join is executed between query points and their candidate neighbors in $S1$ only, the non-neighbors of $q$ in $S1$ belonging to $c1, c4$ and $c6$ are pruned out.
\end{example}

\section{Experimental Evaluation}
\label{sec:experiments}

\subsection{Streams and Environment}
For GeoFlink evaluation, Microsoft T-Drive data \cite{TDriveData} is used, containing the GPS trajectories of 10,357 taxis during the period of February 2 to 8, 2008 in the Beijing city. The total number of tuples in the dataset is 17 million and the total distance of the trajectories is around 9 million kilometres. Each tuple consists of a taxi id, datetime, longitude and latitude. The dataset is loaded into Apache Kafka \cite{KafkaStreaming} and is supplied as a distributed stream to GeoFlink cluster.

For the experiments, a four nodes Apache Flink cluster with GeoFlink (1 Job Manager and 3 Task Managers (30 task slots)) and a three nodes Apache Kafka cluster (1 Zookeeper and 2 Broker Nodes) are used. The clusters are deployed on AIST AAIC cloud \cite{AAIC}, where each VM has 128 GB memory and 20 CPU cores where each core uses Intel skylake 1800 MHz processor. All the VMs are operated by Ubuntu 16.04. 


\subsection{Evaluation}
This section compares our proposed grid-based spatial queries with their respective naive approaches. To keep the comparison fair, efforts are made to distribute the data streams uniformly across the cluster nodes for the naive approaches. The evaluation is presented in terms of system throughput (maximum number of stream tuples processed by the system per second). Unless otherwise stated, following default parameter values are used in the experiments: grid size ($m$): 150 x 150 cells, $r$: 400 meters, $W_n$: 10 seconds, $W_s$: 5 seconds and $k$: 10. Each experiment is performed three times and their average values are reported in the graphs. Since the T-Drive data stream is from Beijing city, we made use of the following rectangular bounding box of the city in terms of longitudes and latitudes for the grid construction: bottom-left = 115.5, 39.6, top-right = 117.6, 41.1. Euclidean distance is used for the distance computation.



Fig. \ref{fig:graphsRangeQuery} evaluates the spatial range query. The throughput of the grid-based approach is far higher compared to the naive approach for all the parameters' variation, mainly due to the effective gird-based pruning. In Fig. \ref{fig:rangeGridSize}, the throughput of the grid-based approach is slightly lower for $m =$ 50x50. This is because at this $m$, individual cells are quite large, resulting in poor pruning. In Fig. \ref{fig:rangeQueryRadius}, we varied query radius ($r$). Since the increase in $r$ results in bigger query result-set, throughput decreases with the increase in $r$. In Figs. \ref{fig:rangeWinSize} and \ref{fig:rangeWinSlide}, window size ($W_n$) and slide step ($W_s$) are varied, respectively. Increasing $W_n$ results in a decrease in the throughput which is quite obvious. On the other hand, increasing $W_s$ in Fig. \ref{fig:rangeWinSlide} results in an increase in the system throughput, because larger slide step means less overlapping as the window slides. This results in the decrease in the number of distance computations and hence increase in the system throughput. Note that the parameters variation do not have much impact on grid-based approach. Please understand that parameters variation have an effect on number of distance computations, query output size and/or query output frequency. Due to the strong pruning, grid-based approach is left with a fraction of distance computations, hence, this effect is not significant in grid-based approach. However, the effects of output size and frequency are same on both the approaches.


Fig. \ref{fig:graphskNNQuery} evaluates the $k$NN query. The throughput of the grid-based approach is almost twice compared to the respective naive approach for most of the variation of the parameters, due to the reasons discussed in the last paragraph. The variation of the different parameters has more or less same effect on the processing of the $k$NN query as in the case of the range query. The only different parameter in the $k$NN query is $k$ (Fig. \ref{fig:kNNk}). Increasing $k$ very slightly decreases the throughput because for the larger $k$ values, larger sorted $k$NN lists need to be maintained.


Fig. \ref{fig:graphsJoinQuery} evaluates the spatial join query. The throughput of the grid-based join query is comparatively far higher than the naive approach, and in most cases more than double. This is because the grid-based approach is capable of pruning a large number of non-neighbors and hence require far less distance computations compared to the naive approach. The trends in the variation of the different parameters are essentially the same as that of the previous queries. The Fig. \ref{fig:graphsJoinQuery} includes variation in the query stream arrival rate as an additional evaluation, the increase of which results in the reduced system throughput, which is obvious. Because with the increase in the number of query points, more computations are needed. However, this reduction is not significant in the grid-based approach, proving the effectiveness of the proposed approach.

\section{Conclusion and Future Work}
\label{sec:conclusion}
This work presents GeoFlink which extends Apache Flink to support spatial data types, index and continuous queries. To enable efficient processing of continuous spatial queries and for the effective data distribution among Flink cluster nodes, a gird-based index is introduced. The grid index enables the pruning of the spatial objects which cannot be part of a spatial query result and thus guarantees efficient query processing. Similarly it helps in uniform data distribution across distributed cluster nodes. GeoFlink currently supports spatial range, spatial $k$NN and spatial join queries on geometrical point objects. Extensive experimental study proves that GeoFlink is quite effective for the spatial queries compared to ordinary Flink. As a future direction, we are working on GeoFlink's extension to support line and polygon data types and other complex query operators. Furthermore, we are looking into other efficient spatial index structures for spatial stream processing.

\begin{acks}
This research was partly supported by JSPS KAKENHI Grant Number JP20K19806 and a project commissioned by the New Energy and Industrial Technology Development Organization (NEDO).
\end{acks}

\bibliographystyle{ACM-Reference-Format}
\bibliography{ref}

\end{document}